# Deposition and Extension Approach to Find Longest Common Subsequence for Multiple Sequences

Kang Ning
University of Michigan, Ann Arbor, MI, USA
kning@umich.edu

**Abstract**
The problem of finding the longest common subsequence (LCS) for a set of sequences is a very interesting and challenging problem in computer science. This problem is NP-complete, but because of its importance, many heuristic algorithms have been proposed, such as Long Run algorithm and Expansion algorithm.

However, the performance of many current heuristic algorithms deteriorates fast when the number of sequences and sequence length increase. In this paper, we have proposed a post process heuristic algorithm for the LCS problem, the Deposition and Extension algorithm (DEA). This algorithm first generates common subsequence by the process of sequences deposition, and then extends this common subsequence. The algorithm is proven to generate *Common Subsequences* (*CSs*) with guaranteed lengths. The experiments show that the results of DEA algorithm are better than those of Long Run and Expansion algorithm, especially on many long sequences. The algorithm also has superior efficiency both in time and space.

## 1. Introduction

The problem of finding the Longest Common Subsequences (LCS) has many applications in different areas of computer science, such as data compression, pattern recognition, file comparison and biological sequence comparisons and analysis (1,2). The LCS of a set of sequences can be formulated as this. For two sequences $S=s_1...s_m$ and $T=t_1...t_n$, $S$ is the subsequence of $T$ ($T$ is the supersequence of $S$) if for some $1 \leq i_1 < ... < i_m \leq n$, $s_j = t_{i_j}$. Given a finite set of sequences $S=\{S_1,S_2,...,S_k\}$, a Common Subsequence (CS) of $S$ is the sequence $T$ such that each sequence in S is a supersequence of $T$, and a LCS of $S$ is the longest possible $T$ among all of CS for this set of sequences $S$.

The LCS problem has been examined extensively by many researchers (refer to (3) for details). The LCS of two sequences can be computed by dynamic programming in $O(n^2)$ time and $O(n^2)$ space, and there are a number of papers on this problem using dynamic programming with reduced time and space (1,4).

Unfortunately, the LCS problem on arbitrary $k$ sequences is a well-know NP-hard problem that is even hard to approximate in the worst case (5). Though there are efficient dynamic programming algorithms on computation of LCS for small $k$ (6,7), these algorithms are not suitable for dataset in which there are many long sequences (2).

Though this problem is NP-hard, it is so important in application that many heuristic algorithms have been proposed to solve the LCS problem (3,5,8). These algorithms compute the common subsequences (not necessarily the longest) of the input sequences. However, their performance (result quality and efficiency) deteriorates fast on large LCS instances. By *large LCS instances*, we mean instances where (a) the sequences in *S* are

*long* (*n* is 100 and more) and (b) there are *many* sequences (*k* is 100 or more). And other instances are *small LCS instances*.

In this paper, we have proposed a deposition and post process approach, the Eeposition and Extension algorithm (DEA), for the longest common subsequence (LCS) problem on a set of sequences. The algorithm that we have proposed is suitable for both small and large LCS instances, and it is effective in time and space even on very large datasets with many long sequences. We have proven the guaranteed performance of the algorithm, analyzed it results empirically, and also empirically showed the superiority of the algorithm to other heuristic algorithms, especially on many long sequences.

2. **Literature review and our approach**

The LCS problem is extensively investigated by many researchers (refer to (2,3)). Though there are algorithms to compute LCS for a small set of sequences (4,9), the optimal solution for this problem on arbitrary number of sequences with arbitrary lengths is NP-hard (5), and there are many heuristic algorithms for the problem.

In the following part, we will use *n* as the length of the sequence, *k* as the number of sequence, and $\Sigma$ as the alphabet set. For the sequences S over alphabet set $\Sigma = \{\sigma_1, \sigma_2, \ldots \sigma_{|\Sigma|}\}$, the **alphabet content** $r_i$ of $\sigma_i$ is defined as the number of $\sigma_i$ in all of the sequences, over the total lengths of all sequences.

**2.1 The Dynamic programming algorithms**

There exists a dynamic programming algorithm to compute the LCS of a set *S* of *k* sequences (9), but it requires $O(n^k)$ time and space, where *n* is the length of the longest sequence in *S*. Hence this algorithm is feasible only for small values of *n* and *k* (6,7). An improvement of such algorithm based on the Four Russians' technique (10) is possible and the time complexity would be $O(n^k/\log n)$, but the hidden constants would not make such an algorithm more appealing than the one in (9) for arbitrary sequences datasets.

**2.2 Heuristic algorithms**

*The Long Run algorithm:* The simple and fast Long Run algorithm is proposed by Jiang and Li (5). For *k* sequences in $S = \{s_1, s_2, \ldots, s_k\}$ on the finite alphabet set $\Sigma = \{\sigma_1, \sigma_2, \ldots, \sigma_{|\Sigma|}\}$. Long Run algorithm finds maximum *m* such that there is $\sigma_i$ in $\Sigma$, and $\sigma_i^m$ is a common subsequence of all input sequences. It outputs $\sigma_i^m$ as the result of LCS. The time complexity of the long run algorithm is $O(kn)$.

*The Greedy algorithm and Tournament algorithm:* These are two variations of an iterative scheme based on combining "best" sequence pairs. Given any pair of sequences, $S_i$ and $S_j$, an optimal subsequence of the pair, LCS($S_i$,$S_j$), can be computed in $O(n^2)$ using dynamic programming. The Greedy algorithm first chooses the "best" sequence pair – the pair that gives the longest LCS($S_i$,$S_j$), Let's call them $S_1$ and $S_2$. The algorithm then replaces the two sequences $S_1$ and $S_2$ by their subsequence, LCS($S_1$,$S_2$). The algorithm proceeds recursively. Thus, we can express it as follows:

$$\text{Greedy}(S_1, S_2, \ldots, S_k) = \text{Greedy}(\text{LCS}(S_1, S_2), S_3, \ldots, S_k) \tag{1}$$

The Tournament algorithm is similar to the Greedy algorithm. It builds a "tournament" based on finding *multiple* best pairs at each round and can be expressed schematically as follows:

$$\begin{aligned}&\text{Tournament}(S_1, S_2, \ldots, S_k) = \\ &\text{Tournament}((\text{LCS}(S_1, S_2), \text{LCS}(S_3, S_4), \ldots, \text{LCS}(S_{k-1}, S_k))\end{aligned} \tag{2}$$

Both Greedy and Tournament algorithms have $O(k^2 n^2)$ time complexity and $O(kn + n^2)$ space complexity.

***The Expansion algorithm:*** The Expansion Algorithm is proposed by Bonizzoni et al. (8). The strategy of this algorithm is to reduce all sequences in set $S$ to streams first, where a stream is the sequence without consecutive identical alphabets. For example, we can reduce the sequence `s1= aaagccttt` to `agct`. Then find all short common streams for sequences, whose lengths are not more than 2. Among them, choose a longest common stream $z$ of all sequences in $S$. Then one can expand all substrings of stream $z$ to find a common subsequence of $S$ with maximal length. For example, the sequence `aag` can be expanded from `ag`. The long run algorithm is also embedded in the expansion algorithm by finding all short common streams and expanding them. The time complexity of the Expansion algorithm is $O(kn^4 \log n)$. To our best knowledge, the Expansion Algorithm is currently one of the best algorithms to the LCS problem on datasets with many sequences.

## 2.3 Other algorithms

The Best Next heuristic algorithm is a typical example of fast LCS heuristic algorithm for two sequences (11,12), and it can be easily extended to a heuristic program for $k$ sequences. Its complexity is $O(|\Sigma|kn)$, where $|\Sigma|$ is the size of alphabets $\Sigma$, $k$ and $n$ are the number and length in $S$, respectively. The machine learning algorithms (13) are also investigated by researchers, with satisfactory results on the computation of two sequences and improved efficiency. Some other algorithms investigate the efficiency of the algorithm on LCS problem (2), or LCS problem on specific sequences (14).

## 2.4 Performance Ratios and Upper Bounds Calculation

To compare the performance of different heuristic algorithms, a performance ratio measurement is proposed. The performance ratio of a heuristic algorithm $A$ over an sequence set S is the value $R_A(S)$, where

$$R_A(S) = |opt(S)| / |CS_A(S)| \tag{3}$$

in which *opt(S)* is the optimal LCS, and $CS_A(S)$ is the result of algorithm $A$, that is, *CS* generated by algorithm $A$. The closer performance ratio is to 1, the better the heuristic.

Long Run (5) and Expansion Algorithm (8) have guaranteed performance ratio $|\Sigma|$, where $|\Sigma|$ is the size of the alphabet set $\Sigma$. Actually, some researchers pointed out that (15) any heuristic algorithm which uses only global information, such as number of symbol occurrences, might return a common subsequence as short as $1/|\Sigma|$ of the length of the optimal results even for two sequences. It was obvious that both Greedy and Tournament do not have approximation ratios.

For large LCS instances, *opt(S)* is generally not available, so that estimation of the upper bound or lower bounds of *opt(S)* is important for calculation of performance ratios.

***Upper Bounds:*** Given a set S of *n* sequences with max sequence length *k*, and the alphabet set $\Sigma = \{\sigma_1, \sigma_2,...\sigma_{|\Sigma|}\}$. The algorithm first choose number $|\Sigma|$ of sequences $\{S_1, S_2,...S_{|\Sigma|}\}$ from the sequences set S, so that $s_i$ is the sequence with most number of $\sigma_i$. Then dynamic programming is applied on $\{S_1, S_2,...S_{|\Sigma|}\}$, and the upper bound of the LCS length is obtained.

If $|\Sigma|$ is large, or the length of the sequences is very long; then the number of sequences chosen can be smaller than $|\Sigma|$. These sequences are chosen in the same manner as descried above, but only part of the alphabets in $\Sigma$ are considered, subject to the alphabet contents of these sequences.

## 2.7 Issues less Addressed and Our Approach

Though there are many heuristic algorithms for the LCS problem (3,5,8), most of these algorithms concentrate on the sequences set that has few long sequences or many short sequences, and few algorithms (8) concentrate on large LCS instances that have many long sequences. However, large LCS instances are common in many applications.

Therefore, it is important to devise an effective algorithm that works well on large LCS instances. In this project, we have focused on large LCS instances. We have developed an effective heuristic algorithm for this purpose, proved its guaranteed performance ratio, and empirically analyzed and compared the performance of the algorithms on simulated and real sequences.

In the following part, we have first described our Eeposition and Extension algorithm (DEA) for the LCS problem, and analyzed some key values. We have then applied our algorithm on simulated and real sequences with different number of alphabets, analyzed and compared our algorithm with other algorithms in the experiments.

## 3. Deposition and Extension Method

In this section, we have proposed the Deposition and Extension algorithm (DEA). In this algorithm, we first generate a common subsequence for a set of sequences by deposition method, and then extend this common subsequence to get common subsequence $CS_{DEA}$.

## 3.1 Deposition Process

A common subsequence *T* (or *template*) is found out through the process of deposition (refer to (19) for application of deposition on SCS problem). In the deposition process, for each of the sequences $S_i$ in *K* sequences, the *front* in deposition step $d_j$ is the number of characters $f_i(d_j)$ that are already processed. At the beginning of the process, every sequence has front 0. For each of the deposition step $d_j$ for sequence $S_i$, the *change* of front is $f_i(d_j)$- $f_i(d_{j-1})$, that is, the difference of current front and the previous front. We also define **search range L**, which limit the number of characters in the sequences that we search for common alphabets. For example, search for alphabet $A_p$ in $S_i$ at deposition step $d_j$ means that we exam the existence of $A_p$ in substring $S_i[f_i(d_j).. f_i(d_j)+L]$. It is easy to see that deposition method without any search range is equivalent to the deposition method with search range L = n.

*MF Deposition Method:* The MF deposition method is based on simple intuition of deposition. In this method, at each deposition step $d_j$, the alphabet $A_j$ which appears in most of the front positions is selected. For a sequence $S_i = a_{i_1} a_{i2}...a_{i_n}$, $S_i[p..q]$ (p≤q) denotes the substring $a_{i_p}...a_{i_q}$. The presence of $A_j$ in every sequence is checked within search range L; that is, for every sequences $s_i$, check if $A_j$ is in the substring $S_i[f_i(d_j).. f_i(d_j)+L]$. If $A_j$ is present in every substring, then each sequence now has the front at position where $A_j$ appears in search range. Define "||" as concatenation, then the template T is extended by $A_j$, $T=T||A_j$. Otherwise, $A_j$ is deposited for those sequences with it as front, but the template is not extended. We do this for multiple iterations, until one or more sequences are completely deposited. Since it chooses the alphabet that appears in most of the fronts, this deposition method is called the "Most Front" (MF) method. Figure 1 gives an example of such process.

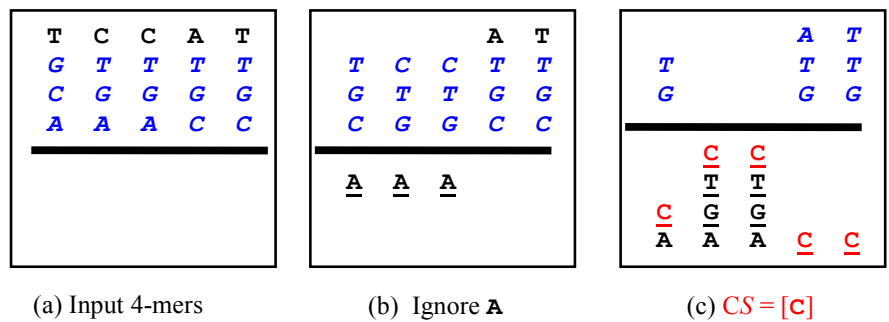

(a) Input 4-mers  (b) Ignore **A**  (c) *CS* = [**C**]

**Figure 1: The step by step illustration of the process of finding long template by deposition. The search range L is set to be 3. (a) is the state before deposition, (b) is the state after first deposition step, and (c) is the state after last step. The result is template *T* = "C", which is NOT the optimal result. Underline indicates the characters deposited, bold bars indicate the fronts, and italics are characters in search range.**

In Figure 1(a), there are five sequences with length 4 each. In Figure 1(b), we have selected character "A" since this alphabet is most frequent on the front ("A, A, A, C, C"), but it is not in every sequence within search range. Next in Figure 1(c), we have tested character "C", and the check returns the positive answer, so the common subsequence *T* is currently "C". Since this "C" is the last character for the second sequence, the process terminates. The template *T*="C".

```
MF Deposition
Input: Seqs set S={S_1, S_2, … S_n}, with maximum length of seq n
       Search range L
Output: A common subsequence T of all the seqs in S

T = "";
While no front f_i(d_j) is equal to seq length
    presence = YES;
    Get alphabet A_j that appears most at fronts of every seqs;
    For every seq S_i in S
            If A_j not in S_i[f_i(d_j), f_i(d_j)+L]
                    presence = NO;
                    Break;
            EndIf
    EndFor
    If presence = = YES
            T = T || A_j;
            For every seq S_i in S
                    f_i(d_{j+1}) = index (S_i, f_i(d_j) + 1, A_j);
            EndFor
    Else
            For every seq S_i in S
                    If front(S_i) = = A_j
                            f_i(d_{j+1}) = f_i(d_j) + 1;
                    EndIf
            EndFor
    EndIf
EndWhile
```

**Figure 2. The pseudo codes for the MF deposition method.**

***MC Deposition Method:*** Another deposition method is to locally minimize the total changes of the fronts. At each step $d_j$, suppose the front is $f_i(d_j)$ for sequence $s_i$, and the search range is L. For every alphabet $A_p$ that is present in every substring (search range) $S_i[f_i(d_j)..f_i(d_j)+L]$, check the total changes of fronts by deposition of this alphabet; that is, check $\sum_{1\leq i\leq k}(f_i(d_{j+1})- f_i(d_j))$ by $A_p$, and select $A_j$ that minimize this value for deposition. Template $T$ is then extended by $A_j$, $T=T||A_j$. If there is no presence of such alphabet $A_j$, then just deposit the alphabet $A_j$ which causes minimum changes of the fronts, and go to next deposition step. Since it chooses the alphabet that locally minimize the total changes, this deposition method is called the "Min Change" (MC) method. An illustration of such process is in Figure 3.

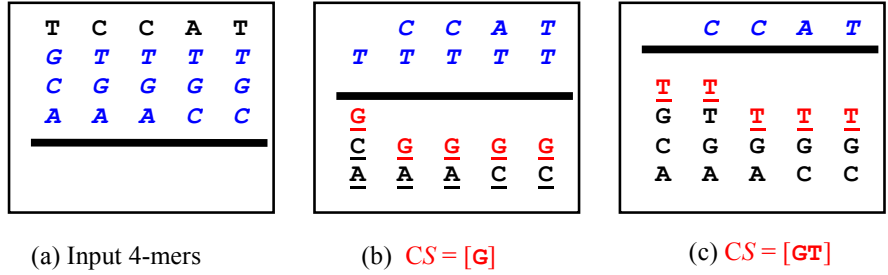

|  (a) Input 4-mers | (b) CS = [G] | (c) CS = [GT] |

**Figure 3. The step by step illustration of the process of finding long template by deposition using the locally minimize the total changes of the front strategy. The search range L is also set to be 3. (a) is the state before deposition, (b) is the state after first deposition step, and (c) is the state after last step. The result is template *T* = "GT", which is the optimal result. Underline indicates the characters deposited, bold bars indicate the fronts, and italics are characters in search range.**

In Figure 3, the same sequences are used as in Figure 1. In Figure 3(a), there are five sequences with length 4 each. In Figure 3(b), the only alphabet in every search ranges is "G", so the template *T* is currently "G". In Figure 3(c), the only alphabet in every search ranges is "T", so the template *T* is currently "GT". Since this "T" is the last character for the first sequence, the process terminates. The template *T*="GT". Compared with Figure 1, we see that MC deposition method can find longer template than MF method o the same example. However, the MC method is not guaranteed to outperform the MF method.

**MC Deposition**
**Input: Seqs set S={$S_1$, $S_2$, … $S_n$}, with maximum length of seq k**
     **Search range L**
**Output: A common subsequence T of all the seqs in S**

T = "";
**While** no front $f_i(d_j)$ is equal to seq length
    presence = YES;
    **For** all of the alphabet $A_p$ that appears in range [$f_i(d_j)$, $f_i(d_j)$+L] for at least one seqs $S_i$
        Find $A_j$ that minimize $\sum_{1\leq i\leq k}(f_i(d_{j+1})- f_i(d_j))$
    **EndFor**
    **For** every seq $S_i$ in S
        **If** $A_j$ not in $S_i$[$f_i(d_j)$, $f_i(d_j)$+L]
            presence = NO;
            **Break**;
        **EndIf**
    **EndFor**
    **If** presence = = YES
        T = T || $A_j$;
        **For** every seq $S_i$ in S
            $f_i(d_{j+1})$ = index ($S_i$, $f_i(d_j)$ + 1, $A_j$);
        **EndFor**

```
    Else
        For every seq S_i in S
            If front(S_i) == A_j
                f_i(d_{j+1}) = f_i(d_j) + 1;
            EndIf
        EndFor
    EndIf
EndWhile
```

Figure 4. The pseudo codes for the MC deposition method.

***Template Pool:*** The template pool is important for the performance of any heuristic algorithms on the LCS problem. This is because a single good template generated by deposition process is not guaranteed to be extended to a good *CS* result. Since a template pool contains many templates for extension, the chance of getting good results is much higher.

We have used template pools for our Deposition and Extension algorithm. The pool contains the template generated by the Deposition process, and the "basic" templates each of which is a single alphabet in $\sum$. For example, for 4-alphabet sequences composed of alphabets $\{\alpha_1, \alpha_2, \alpha_3, \alpha_4\}$, each of these alphabets is one "basic" template in the pool. These templates are then extended by the Extension process, and the longest *CS* result is obtained. Note that by using "basic" templates and extension method, DEA's results are at least as long as Long Run's results on the same datasets (details in Extension Process section).

Intuitively, the larger the $|\sum|$, the better relative performance of the template-based algorithms, compared with other alignment-based algorithms. This is because the template pool based algorithms have provided more flexibility for the extensions; thus can generate longer *CS* results. Based on these thought, we think that Deposition and Extension algorithm can outperform Long Run a lot if the size of $\sum$ is large. Experiments results confirmed this expectation.

### 3.2 Extension Process

After the template is obtained by Deposition process, the Extension process can be applied on template pool to get the *CS* result. This is the post process that would yield improved heuristic results.

In the Extension process, for each of the templates, (a) every alphabet is first added at either end of *template* one at a time, and tests if the extended sequence is the common subsequence of all the sequences in the set. The process continues until no more character can be added at either end of the common subsequence. (b) Then the template is extended from within. In this step, for each of the character *a* in the current template, we try to replace *a* with $a^q$, such that $a^q$ is a string with $q$ ($\geq 2$) number of *a*, and tests if the extended sequence is the common subsequence of all the sequences in the set. The process continues until common subsequence can not be extended.

Both of deposition and extension processes are very important to DEA algorithm. The deposition process has generated good template, which is essential to the good *CS* result.

And the good extension process fully exploits the power that a good template can offer. Together, they can produce good *CS* results.

The time complexity of the MF method is $O(k^2n|\Sigma|)$, and the time complexity of MC is also $O(k^2n|\Sigma|)$. The extension method has the time complexity of $O(k^2n|\Sigma|)$. So the for both of MF method and MC method, the total time complexity of the Deposition and Extension algorithm is $O(k^2n|\Sigma|)$. This is larger than the time complexity of the Long Run algorithm, but much less that that of the Expansion Algorithm.

### 3.3 Guaranteed Performance Ratio

Our Deposition and Extension Algorithm (using template pools) also have a guaranteed performance ratio of $|\Sigma|$, since by using the template pool, the *CS* generated by our algorithm is at least as long as the ones generated by Long Run algorithm. However, our experiment results have shown that the performance ratios of DEA are much lower than the guaranteed performance ratio $|\Sigma|$.

### 3.4 Analysis of Search Range

As described above, a good value of search range L directly affect the performance of the deposition process (both for MF and MC), and it can be evaluated by analysis of the sequences. In the following part, we will give the upper and lower bounds for the search range L, and then determine L for our algorithm.

***Empirical Analysis:*** For estimation of the approximate values of search range *L*, suppose the length of each sequence is *n*, and the expected length of LCS is *E(|LCS|)*, then it is easy to see that

$$L \geq \left\lceil \frac{n}{E(|LCS|)} \right\rceil \qquad (5)$$

For which only the upper bound of *E(|LCS|)* could be computed, which would correspond to lower bound of search range L. It is observed that for too short L, it is not likely to generate reasonably long templates. This means that the values of L should be at least 2 for $|\Sigma|=4$, at least 3 for $|\Sigma|=20$, and at least 6 for $|\Sigma|=100$. These lower bounds of search range L are also listed in Table 1.

| Sequence length (N) | $|\Sigma|$ | E(|LCS|) (upper bounds) | L (lower bounds) |
| --- | --- | --- | --- |
| 25 | 4 | 15.86 | 2 |
|  | 20 | 8.75 | 3 |
|  | 100 | 4.44 | 6 |
| 100 | 4 | 65.24 | 2 |
|  | 20 | 34.83 | 3 |
|  | 100 | 18.05 | 6 |
| 500 | 4 | 326.01 | 2 |
|  | 20 | 180.56 | 3 |
|  | 100 | 89.81 | 6 |
| 5000 | 4 | 3268.26 | 2 |
|  | 20 | 1819.85 | 3 |
|  | 100 | 900.19 | 6 |

**Table 1: The upper bounds of the expected length of LCS on sequences on different $\sum$ based on extensive simulation. And the lower bounds of the search range values. Notice again that the upper bounds are very loosely defined just to make it correct.**

Another important factor is the probability that any alphabet $\sigma_i$ appears in the search ranges of all the sequences. It is easy to see that for random sequences, if search range $L$ is larger than $|\sum|$, then the probability is high.

Formally, for a certain alphabet $\sigma_i$ in $\sum$ which represents $r_i*kn$ characters in the sequences, the *existence probability P*, which represent the probability that it appears in the search range for all of the $k$ sequences (each of length at most $n$) is

$$P = (1-(1-r_i)^L)^K \quad (6)$$

which is independent from the length of the sequences. Suppose we have analyzed part of the sequences and get $r_i$, and we are given the existence probability $P$, then the search range can be computed as

$$L = \frac{\log(1-P^{\frac{1}{K}})}{\log(1-r_i)} \quad (7)$$

We have computed the change of L with different $r_i$ (gamma) and $P$. Results are shown in Figure 5.

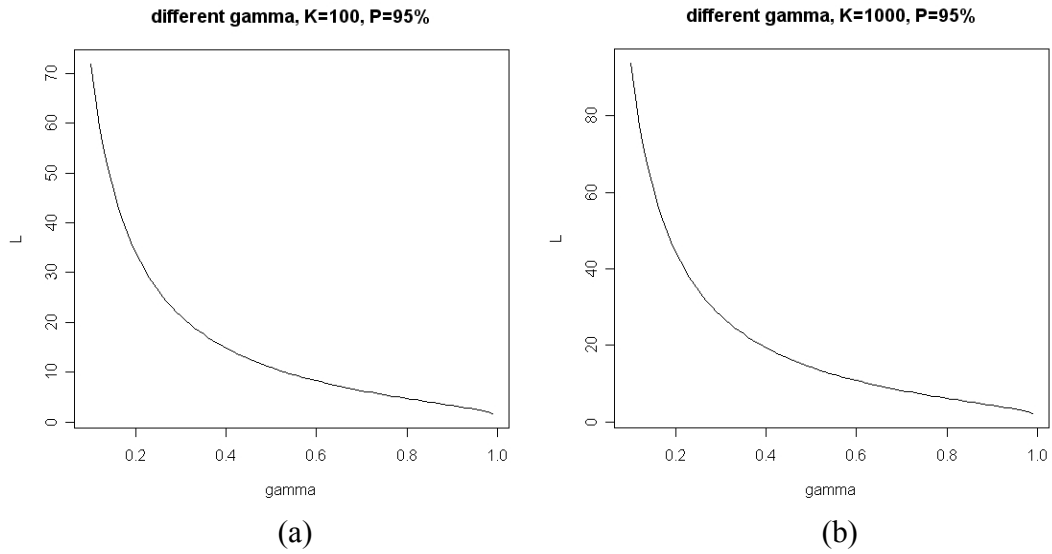

(a)          (b)

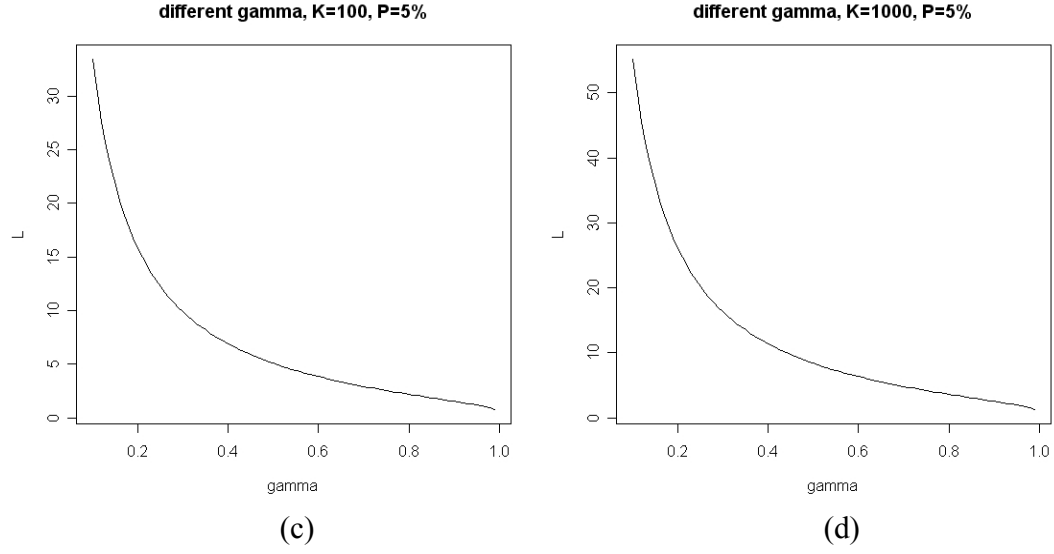

(c)                                       (d)

**Figure 5. The relationship between L and $r_i$. (a) and (b) are for P=95%, with 100 and 1000 sequences respectively; (c) and (d) are for P=5%, with 100 and 1000 sequences respectively.**

From Figure 5 (a) and (b), we see that to achieve P=95%, L>30 for $r_i$ = 0.2, L≥11 for $r_i$ = 0.5 and L≥5 for $r_i$ = 0.8 are required for alphabet $\sigma_i$. L should also be longer for more sequences to achieve higher value of P. For $r_i$ over 0.5, there is not much difference in search range for $K$=100 and 1000. On the other hand, From Figure 5 (c) and (d), we observe that if the search range is too short, such as L≤5 for $r_i$ = 0.5 and L≤3 for $r_i$ = 0.8, then it is unlikely (P = 5%) that there exist such common alphabet $\sigma_i$ within search rage for every sequences, even for only K=100 sequences.

As a special case, suppose all of the alphabets are evenly distributed in the sequences, then the existence probability $P$ is

$$P = (1 - (\frac{|\Sigma|-1}{|\Sigma|})^L)^K \qquad (8)$$

Given the specified value of P, as well as K and $|\Sigma|$, the search range can be computed as

$$L = \frac{\log(1 - P^{\frac{1}{K}})}{\log(\frac{|\Sigma|-1}{|\Sigma|})} \qquad (9)$$

So that we can compute the value of L with specified existence probability P and number of sequences K. In Figure 6, we have illustrated the relationship between search range (L) and the number of sequences (K), given specific alphabet size $|\Sigma|$ and probability (P) that a given alphabet $\sigma_i$ appears in the search range L.

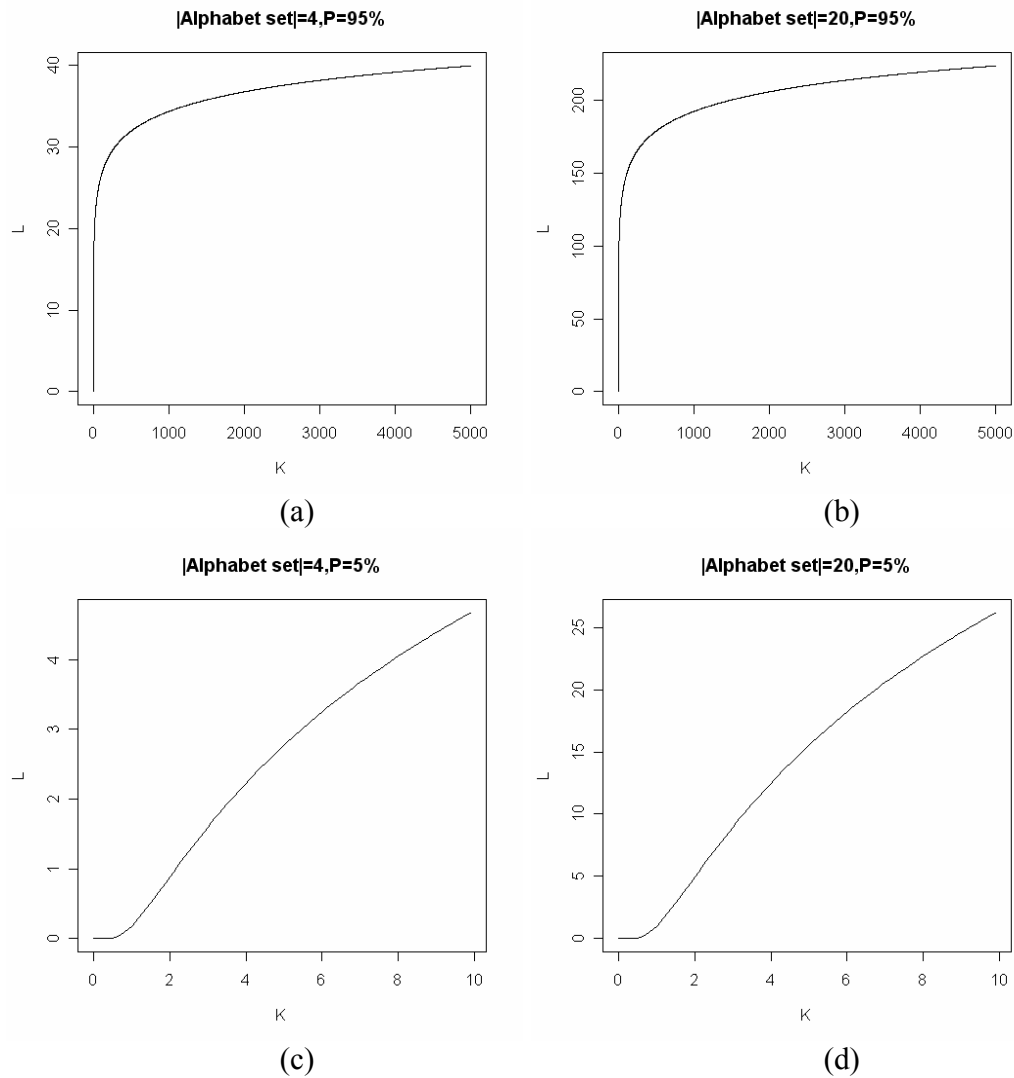

**Figure 6: The relationship between L and K given existence probability P that a given alphabet appears in the search range. (a) and (b) are for P=95%, (c) and (d) are for P=5%.**

From the relationship between search range L and the number of sequences K, we have observed from Figure 5 (a) and (b) that we can confidently (P = 95%) find a specified common alphabet in all of the sequences within search range L=40 for $|\sum|=4$ and as many as 5,000 sequences; and within search range L=165 for $|\sum|=20$ and as many as 250 sequences. On the other hand, from Figure 5 (c) and (d), we observe that if search range is short, such as L=4 for $|\sum|=4$, and L= 25 for $|\sum|=20$, then it is unlikely (P = 5%) that there exist one specified common character within search rage for all sequences, even for as few as K=10 sequences. These results are consistent with those in Figure 5.

Intuitively, on a set of random sequences, if we can confidently (P=95%) find a common alphabet in a search range L, then the expected length of LCS is (N-L)/(L/2)+1, which can be simplified as

$$(2N/L)-1 \qquad (10)$$

This is because on a random set of sequences, the alphabets are distributed evenly in the sequences. In each step that a common alphabet is found, the average "deposition progress" of a sequence is L/2; except for the last step, in which the "deposition progress" of L will end the whole process. This estimation of the length of LCS is consistent with those bounds given by Table 1. Moreover, this estimation is in very good consistance with the actual results on simulated and real sequences (in Experiments), especially on dataset with many sequences ($\geq$5,000).

Based on above analysis, we have chosen appropriate values of search range L for DEA algorithm. Specifically, we have used L = *min(50, n/i)* (*n* is the length of sequence) as the value of search range to adapt the different sequences to get good results. Value 50 is chosen because based on results in Figure 6, given L=50, and alphabet content $r_i \geq$ 20%, the number of the probability of one alphabet presents in all search ranges is above 95%, up to as many as K=1,000 sequences. Integer *i* is chosen so that *i = argmax(L= min(50, n/i) result in longest CS) | i*$\in$[1,10]. By this way, at least 10% of the possible search range values are examined. In the following part, we will see the importance of using the search range.

## 4. Experiments

### 4.1 Experiment Settings

The Deposition and Extension (DEA) algorithm is written in Perl. The experiments are performed on a PC with 3 GHz CPU and 1 GB memory, running a Linux system.

We have selected Long Run (LR) (5) and Expansion Algorithm (EA) (8), the two algorithms with guaranteed performance ratios for comparison with our algorithm. For the Expansion Algorithm, if not specified, the number of runs is set to be 4 (the reason for this setting will be explained later). For DEA, the search range L is set as described above. Both MF and MC methods are used for performance analysis.

For the computation of the performance ratios, we have computed upper bounds of the CS results on large LCS instances. Note that the performance ratios are only the upper bounds of the performance ratios by this way.

There are two kinds of datasets: simulated and real. Simulated datasets contain simulated 4-alphabet sequences, and simulated 26-alphabet sequences. For the simulated datasets, if not specified, there are 10 set of randomly generated sequences for each settings, and the results are averaged out. It is important to note that each of the alphabets has different (random) occurrence probability.

The real datasets are more meaningful than simulated datasets for real applications. The real 4-alphabet sequences (DNA sequences) are from NCBI viral genomes (http://www.ncbi.nlm.nih.gov/genomes/VIRUSES/viruses.html) (randomly selected from rival1 genomes), and the real 20-alphabet sequences (protein sequences) are from SwissProt (Swiss-Prot Release 45.5 of 04-Jan-2005, http://us.expasy.org/sprot/)

(randomly selected from SwissProt protein sequences). According to the different settings (sequences length, number of sequences in the datasets, and alphabet content of the sequences) of these datasets, we have grouped them into a few cases (Supplementary Data), and the results are drawn based on the average results of each case.

Real sequences also include real text sequences with equal or more than 26 alphabets. We have selected chain letters and spam mails as real text sequences for analysis. The chain letters are messages that attempt to induce the recipient to make a number of copies of the letter and then pass them on to one or more new recipients. The chain letter sequences that we have used are selected from achieve of chain letters (20) from http://www.cs.uwaterloo.ca/~mli/chain.html. There are 33 chain letters in the dataset, each of then is treated as a text sequence. The average length of these letters is about 2,000 characters. No two letters are identical. Spam mails are nearly identical messages sent to thousands (or millions) of recipients. The spam mails used are obtained from CIAC scam chains "Nigerian 419 Spam" repository at http://hoaxbusters.ciac.org/HB419ScamList.shtml. There are 41 spam mails different in contents, we have only used there contents, and treat each of them as one text sequence (Supplementary Data). The analysis of LCS on both chain letters and spam mails are meaningful, since the analysis of the similarities of chain letter text sequences can help to reveal the evolution relationships among the letters, while the analysis on spam mail texts is a key step in retrieval of the spam features of spam mails.

### 4.2 Results

We have first analyzed the performance of Deposition methods, the Extension method, and the effect of template pool. Then we have compared the results of *CS* among different algorithms – Long Run (5), Expansion Algorithm (8) and DEA algorithm, as well as the upper bound results.

***Analysis on different aspects of the DEA algorithm:*** We have first analyzed the performance of Deposition method and the Extension method for DEA algorithm on simulated datasets. The template pool is used. The results are shown in Table 2.

| No. sequences (k) | Sequence length (n) | MF | MF & Extension | MC | MC & Extension |
|---|---|---|---|---|---|
| **100** | **100** | 24.30 (6.69) | **25.75 (6.76)** | 24.60 (6.23) | **25.65 (6.87)** |
| | **1000** | 306.70 (73.09) | **316.75 (73.17)** | 311.80 (69.76) | **320.60 (70.70)** |
| **500** | **100** | 20.65 (6.58) | **22.20 (6.62)** | 21.10 (6.36) | **22.25 (6.77)** |
| | **1000** | 292.05 (72.12) | **301.10 (72.54)** | 293.65 (70.23) | **301.60 (71.44)** |
| **1000** | **100** | 20.00 (6.54) | **21.30 (6.72)** | 20.25 (6.19) | **21.15 (6.48)** |
| | **1000** | 289.60 (71.54) | **297.25 (72.18)** | 290.75 (70.41) | **297.70 (71.06)** |
| **5000** | **100** | 17.60 (6.18) | **18.70 (6.42)** | 17.85 (6.17) | **18.95 (6.38)** |
| | **1000** | 286.00 (71.55) | **292.25 (71.31)** | 286.50 (71.23) | **292.70 (70.45)** |

**Table 2. Results of the Deposition method and the Extension method for the Deposition and Extension algorithm. In each cell, the "average value (standard deviation)" is given for each setting.**

It can be observed from Table 2 that the MC deposition method can result in longer LCS templates than the MF deposition method, though their differences are little. It can also be observed that the Extension method can effectively increase the length of the results. For sequences of length 100, the increases range from 1 to 2 characters, while the increases are over 10 characters for sequences of length 1000. We also observed that the lengths of resulting *CS* decreases as the number of sequences increase, but not significantly. For sequences with lengths 1000, the results only decreases by about 30 characters as the number of sequences increases from 100 to 5000.

We have then examined the effect of the template pool. We have used the template only generated from MC (min change) deposition process, compared with template only generated by single symbol, as well as DEA results based on template pool. Results are shown in Table 3. Both lengths of template and the final results are given.

| No. sequences (k) | Sequence length (n) | Single Symbol (Template length) | Single Symbol (CS length) | MC (Template length) | MC (CS length) | DEA (Template length) | DEA (CS length) |
|---|---|---|---|---|---|---|---|
| 100 | 100 | 1 | **22.65 (6.33)** | 23.25 (5.88) | **25.65 (6.87)** | 24.60 (6.23) | **25.65 (6.87)** |
|  | 1000 | 1 | **293.75 (70.69)** | 311.80 (69.76) | **320.60 (70.70)** | 311.80 (69.76) | **320.60 (70.70)** |
| 500 | 100 | 1 | **19.15 (7.59)** | 20.15 (6.48) | **22.25 (6.77)** | 21.10 (6.36) | **22.25 (6.77)** |
|  | 1000 | 1 | **287.75 (71.70)** | 293.55 (70.14) | **301.60 (71.44)** | 293.65 (70.23) | **301.60 (71.44)** |
| 1000 | 100 | 1 | **18.05 (7.50)** | 19.60 (6.67) | **21.15 (6.48)** | 20.25 (6.19) | **21.15 (6.48)** |
|  | 1000 | 1 | **286.45 (71.51)** | 288.50 (69.51) | **297.70 (71.06)** | 290.75 (70.41) | **297.70 (71.06)** |
| 5000 | 100 | 1 | **15.50 (7.45)** | 15.25 (8.15) | **18.95 (6.38)** | 17.85 (6.17) | **18.95 (6.38)** |
|  | 1000 | 1 | **284.25 (72.12)** | 283.55 (69.65) | **292.60 (70.36)** | 286.50 (71.23) | **292.60 (70.45)** |

**Table 3. Results on the effect of the template pool. In each cell, the "average value (standard deviation)" is given for each setting.**

From results in Table 3, we see that the template pool is effective. By only using the single symbol as template, exactly as used by the Long Run algorithm, the extension process can extend the lengths of results greatly, but still about 3~10 characters shorter than the results of DEA. For the template only generated from MC deposition method, the Extension process can generally increase the length of template by 2 characters for sequences of length 100, and about 10 characters for sequences of length 1000. In most of these cases, the results are already the same as using the template pools. But for k=5000 and n=1000, we have found that the results by using the template pool is still a little longer than those obtained from template only generated from MC deposition process. Moreover, using template pool only adds a small overhead since extension from single alphabet is very fast.

It is also interesting to see the results for sequences in which certain alphabet has specific alphabet contents. We have used the MF and MC deposition methods for DEA (not using template pool), and tested 4-alphabet sequences in which each alphabet has different alphabet contents. Suppose the alphabets are $\{\alpha_1, \alpha_2, \alpha_3, \alpha_4\}$, we define the probability that the total alphabet contents of $\{\alpha_1, \alpha_2\}$ is $\beta$, and analyze the results given different $\beta$. The results are shown in Table 4. Note that due to symmetric property, analysis on 10%~50% is enough.

| No. sequences (k) | Sequence length (n) | $\beta$ (%) | MF & Extension | MC & Extension |
|---|---|---|---|---|
| 100 | 1000 | 10 | 443.20 (3.97) | 446.00 (3.00) |
| | | 20 | 389.80 (5.55) | 391.20 (4.07) |
| | | 30 | 337.50 (1.86) | 338.70 (4.03) |
| | | 40 | 285.10 (2.30) | 290.00 (1.84) |
| | | 50 | 243.60 (2.11) | 250.00 (3.38) |
| Average | | | 339.84 (71.42) | 343.18 (69.95) |
| 1000 | 1000 | 10 | 419.80 (2.14) | 418.80 (1.54) |
| | | 20 | 369.00 (3.61) | 369.70 (3.98) |
| | | 30 | 320.70 (0.78) | 320.30 (2.24) |
| | | 40 | 272.10 (2.88) | 272.50 (1.91) |
| | | 50 | 224.80 (1.08) | 226.70 (1.42) |
| Average | | | 321.28 (68.90) | 321.30 (68.00) |

**Table 4. The effect of different alphabet contents on the results of algorithm. In each cell, the "average value (standard deviation)" is given for each setting.**

Intuitively, as the values of $\beta$ increase, the variances among sequences will become larger, and the lengths of CS results will become shorter. This is proved in our experiment results. From Table 4, we observe that the different alphabet contents have great effect on results. For datasets with 100 sequences of length 1000 each, the lengths of LCS results decrease by about 100 characters as the values of $\beta$ increase from 10% to 50%. There are similar results for datasets with 1000 sequences. And these show that the different alphabet contents really affect the CS results greatly. For the two different deposition methods, the MC deposition method outperformed the MF deposition method, despite the different alphabet contents.

***Comparisons on simulated sequences:*** We have then compared DEA with Long Run (LR) and Expansion Algorithm (EA) on simulated 4-alphabet sequences. The DEA without search range (that is, L = n) is also examined. The comparison results on simulated datasets are shown in Table 5.

| No. sequences (k) | Sequence length (n) | Upper Bound | LR | EA | DEA (MF) | DEA (MC) | DEA (MC, L=n) |
|---|---|---|---|---|---|---|---|
| **5** | **100** | 56.60 | 29.05 (6.67) | 39.35 (5.06) | **39.60 (6.33)** | **41.15 (5.70)** | **41.00 (5.80)** |
| **10** | **100** | 55.80 | 27.10 (6.50) | 32.75 (6.86) | **34.30 (6.02)** | **35.50 (5.48)** | **34.75 (5.84)** |
| **50** | **100** | 54.35 | 24.55 (5.98) | 25.90 (7.03) | **28.00 (6.96)** | **28.20 (6.85)** | **27.05 (6.94)** |
| **100** | **100** | 53.65 | 22.65 (6.33) | 23.60 (7.21) | **25.75 (6.76)** | **25.65 (6.87)** | **24.85 (6.81)** |
|  | 1000 | 673.35 | 293.75 (70.69) | 308.45 (72.77) | **316.75 (73.17)** | **320.60 (70.70)** | **320.60 (70.70)** |
| **500** | **100** | 51.30 | 19.15 (7.59) | 19.30 (7.74) | **22.20 (6.62)** | **22.25 (6.77)** | **21.70 (6.84)** |
|  | 1000 | 674.45 | 287.75 (71.70) | 291.35 (72.49) | **301.10 (72.54)** | **301.60 (71.44)** | **301.60 (71.44)** |
| **1000** | **100** | 50.70 | 18.05 (7.50) | 18.05 (7.50) | **21.30 (6.72)** | **21.15 (6.48)** | **20.65 (6.79)** |
|  | 1000 | 672.55 | 286.45 (71.51) | 288.40 (71.95) | **297.25 (72.18)** | **297.70 (71.06)** | **297.50 (70.86)** |
| **5000** | **100** | 49.45 | 15.50 (7.45) | 15.50 (7.45) | **18.70 (6.42)** | **18.95 (6.38)** | **18.55 (6.58)** |
|  | 1000 | 672.2 | 284.25 (72.11) | 284.25 (72.12) | **292.25 (71.31)** | **292.70 (70.45)** | **292.70 (70.45)** |

**Table 5: The comparison of *CS* results among different algorithms, and the upper bound results. "Deposition and Extension (L=N)" means that the search range is the whole sequence. The datasets used are simulated sequences. In each cell, the "average value (standard deviation)" is given for each setting.**

Theoretically, the guaranteed performance ratio $R_A(S)$ is 4 for 4-alphabet sequences. The above comparison results show that for the datasets with many sequences, the performance ratios of all these heuristic algorithms are better than the guaranteed performance ratio, but not as good as ratios for the datasets with few sequences, as 1.1 for EA and 1.5 for LR shown in (8). Compared with upper bounds, we can see that these performance ratios are large ($\geq 2$) for datasets with many ($\geq 50$) long ($\geq 500$) sequences. The main reason that the performance ratios are not as good is probably due to large LCS instances.

Compared among these algorithms, DEA (based on both MF and MC for deposition) have comparative or better (in most of the cases) performance ratios than EA, and better performance ratios than LR algorithm. The superiority of the DEA to LR indicates that the templates generated from DEA are generally better than the simple one-alphabet template. For datasets with many long sequences, the performance of DEA is better than that of EA, showing that DEA have better template and extension method on these datasets. The performance of DEA using MC method is slightly better than the algorithm using the MF method, with results about 1 character longer, and smaller standard deviation. This means that DEA using MC method has better and more stable results.

We emphasis that for EA, we have set the number of runs to be 4 in this experiment; on the same sequences datasets, we have also performed the EA with 32 runs. The results (not shown here) show that the increase of the number of runs from 4 to 32 can only increase the lengths of *CS* a little (0~2 characters, depending on the length of the sequences) for small datasets (k≤10, n≤500), and almost no different results for large datasets. So for efficiency consideration, we have used 4 runs for the following experiments.

The use of search range is also important for DEA. In Table 5, we have also used DEA with search range L=n in the MC deposition method to compare with normal DEA based on search range $L = min(50, n/i)$. Table 5 shows that the results based on L=n are not as long as results of DEA with flexible search ranges. This empirically proved our hypothesis, and indicates that the flexible search range is important for the performance of DEA.

For the simulated text sequences based on 26 alphabets, *CS* of sequences is generated by different algorithms, including LR, EA and DEA. The upper bounds are also calculated on the datasets. The results are shown in Table 6.

| No. sequences (k) | Sequence length (n) | Upper Bound | LR | EA | DEA (MF) | DEA (MC) |
|---|---|---|---|---|---|---|
| 100 | 100 | 17.3 | 0.9 (0.30) | 0.9 (0.30) | **0.9 (0.30)** | **0.9 (0.30)** |
| 100 | 1000 | 318.2 | 28.2 (0.60) | 28.2 (0.60) | **31.9 (0.70)** | **32.2 (0.78)** |
| 500 | 1000 | 317.9 | 24.7 (0.64) | 24.7 (0.64) | **26.8 (0.40)** | **27.4 (0.61)** |
| 1000 | 1000 | 319.3 | 23.3 (0.78) | 23.3 (0.78) | **25.8 (0.40)** | **27.5 (0.87)** |

**Table 6: The *CS* of simulated 26-alphabet sequences by different algorithms. In each cell, the "average value (standard deviation)" is given for each setting.**

The upper bound of the performance ratio of LR, EA and DEA is 26. Comapred with upper bound of LCS results in Table 6, we see that the performance ratios of these algorithms are still much lower than the upper bound. The results show that the performance of LR and EA give the same results. We speculate that for such a large alphabet size as 26, the EA can not perform well.

The results also show that DEA outperforms LR and EA. The lengths of the *CS* are always 1 or 2 characters longer than the results of the Long Run and the Expansion Algorithm. The strategy to generate the good templates, as well as the well designed post process strategy, should be the reason for this superiority. The little superiority of MC deposition method over MF deposition method is probably due to its local optimization. The results of DEA (on both MF and MC deposition method) are also very stable.

***Comparisons on real sequences:*** We have also compared these algorithms on real 4-alphabet and 20-alphabet sequences (DNA and protein sequences). For 4-alphabet sequences, we have used 6 cases (groups of datasets) with different number of sequences and sequence lengths. 6 representative cases (groups of datasets) of 20-alphabet sequences are also used for comparison.

The results are shown in Table 7. As can be observed in the results, we have truncated the lengths of the sequences to specific values. These do not affect the performance of the algorithms much, but the measurement of the average and standard deviations are more meaningful.

|  | No. sequences (k) | Sequence length (n) | Upper Bound | LR | EA | DEA (MF) | DEA (MC) |
|---|---|---|---|---|---|---|---|
| **4-alphabet** | | | | | | | |
| **Case1** | 100 | 100 | 36.2 | 12.8 (0.98) | 13.3 (1.19) | **14.9 (1.04)** | **15.8 (0.98)** |
| **Case2** | 100 | 500 | 340.5 | 79.7 (7.27) | 94.1 (7.74) | **98.6 (7.46)** | **94.3 (6.40)** |
| **Case3** | 100 | 1000 | 718.4 | 150.8 (15.95) | 164.0 (15.53) | **198.2 (14.23)** | **191.4 (10.43)** |
| **Case4** | 500 | 100 | 27.1 | 7.6 (1.56) | 7.6 (1.56) | **9.8 (1.47)** | **10.7 (1.00)** |
| **Case5** | 500 | 500 | 326.0 | 50.3 (9.75) | 57.5 (15.22) | **65.8 (15.54)** | **71.0 (7.11)** |
| **Case6** | 500 | 1000 | 784.4 | 107.3 (25.68) | 119.4 (24.79) | **142.9 (19.22)** | **152.1 (22.30)** |
| **20-alphabet** | | | | | | | |
| **Case1** | 100 | 100 | 20.4 | 2.2 (0.60) | 2.3 (0.63) | **2.5 (0.80)** | **2.8 (0.60)** |
| **Case2** | 100 | 500 | 211.8 | 18.4 (4.69) | 18.4 (4.69) | **24.8 (2.71)** | **22.0 (3.44)** |
| **Case3** | 500 | 100 | 19.6 | 0.9 (0.70) | 0.9 (0.70) | **0.9 (0.70)** | **0.9 (0.70)** |
| **Case4** | 500 | 500 | 205.7 | 12.9 (3.62) | 14.2 (4.27) | **18.5 (2.94)** | **13.7 (3.49)** |
| **Case5** | 1000 | 100 | 19.5 | 0.1 (0.3) | 0.1 (0.30) | **0.1 (0.30)** | **0.1 (0.30)** |
| **Case6** | 1000 | 500 | 164.2 | 9.8 (1.99) | 10.4 (2.29) | **14.1 (1.14)** | **11.7 (1.27)** |

**Table 7: The comparison of CS results among different algorithms, and the upper bound results. The datasets used are real 4-alphabet and 20-alphabet sequences. In each cell, the "average value (standard deviation)" is given for each setting.**

The real 4-alphabet datasets contains many long sequences, and DEA outperform LR and EA on these datasets. These show the superiority of DEA on the real 4-alphabet sequences.

Comparing Table 7 with Table 5, we have also observed that for real 4-alphabet sequences, the *CS* generated by LR, EA and DEA are much shorter than those obtained from simulated datasets. The short *CS* is probably due to the high variances (very different alphabet contents) among the sequences in each dataset. For each of the simulated 4-alphabet sequences datasets, the alphabet contents are predefined, so the differences among sequences are not as significant as in real sequences. As the sequences in one real dataset are very different from each other, the *CS* of these sequences is bound to be short. The results of DEA using MC method are similar to the results of MF method; but MC results are much longer when there are many (k=500) sequences.

As for the real 20-alphabet sequences, DEA is also superior to LR and EA on the datasets. For long sequences, the performance ratios are much lower than the guaranteed performance ratios. The comparison of the results on 4-alphabet and 20-alphabet sequences also indicates that $|\sum|$ does not affect the superiority of the DEA to LR and EA. This is probably due to the flexibility of DEA's strategy for generation of template. The results of DEA using MC method are not as good as the results of MF method, probably due to the large alphabet size.

We have also tested these algorithms on the real text sequences, and the results are shown in Table 8. To make the results meaningful, we set a constraint that *CS* can not be purely consecutive non-word.

| Name | No. sequences (k) | Sequence length (n) | No. alphabets | Upper Bound | LR | EA | DEA (MF) | DEA (MC) |
|---|---|---|---|---|---|---|---|---|
| Chain letters | 33 | ~2,000 Min: 1,421 Max: 2,136 | 74 | 995 | 152 | 152 | **159** | 164 |
| Spam mails | 41 | ~3,000 Min: 1,048 Max: 4,563 | 53 | 992 | 88 | 88 | **177** | 179 |

**Table 8: The *CS* of real text sequences by different algorithms.**

From Table 8, we can observe that for chain letters, the result of DEA is slightly better than other algorithms, while for spam mails, the CS length generated by DEA algorithm outperform the results of LR and EA a lot. This is partially due to the longer length of the spam mails, and the different structure of the spam mail as compared to the chain letters may also be the reason. Compared with upper bound of LCS length in Table 8, we observe that the performance ratios of these algorithms are much better than the upper bounds, while DEA has the best performance ratios.

We have checked the template generated by DEA using MF deposition method, and notice a very interesting property. The template generated for the chain contains many meaningful words or phrases such as "with luck", "the original copy", "in new", "it has" "around the world", "nine ties", which are highly related with the contents of the chain letters. The template generated for spam mails contains words or phrases such as "dear sir", "I", "me", "the" and "with". Spammers intentionally make some typos in the mail to avoid different types of filtration, but these typos are less related with the contents, and they are not common words. This analysis shows that the *CS* generated by Deposition and Extension algorithm on real text sequences can reveal parts of the common meaningful information contained in the texts.

*Efficiencies:* Time complexes of LR, RA and DEA algorithms are already analyzed previously. In experiments, DEA (implemented in Perl) is very fast, even for datasets with many (≥1000) long (≥ 1000) sequences (a few minutes). Using MC deposition method is slightly slower than using MF deposition method. Though it is slower than LR (our implementation in Perl), it is many times faster than EA (implemented in C).

Since the processing time for DEA is very short, even for datasets with many long sequences, we have the good opportunity to provide software and web services that can be used for such instances. The software is available upon request.

## 5. Conclusion

The LCS problem is so important that there is constant need for an efficient and accurate algorithm for this problem. In this paper, we have proposed a post process approach, the Deposition and Extension algorithm (DEA) on the LCS problem. And we have focused specifically on large LCS instances.

The DEA algorithm first generates good template by MF or MC deposition method, and then generate a template pool for extension. The algorithm then tries to extend template in the template pool to get good CS results. In the deposition process, the DEA algorithm used a search range for efficient and effective alphabet search. And in the extension process, the DEA algorithm extends the templates both at the two sides and within the template. These strategies are proven empirically to be powerful for solving the LCS problem heuristically.

We have compared the performance of our DEA algorithm with some popular heuristic algorithms (Long Run and Expansion Algorithm) on different sequences. Our DEA algorithm has the guaranteed performance ratios, and in practice have superior performance ratios than Long Run algorithm. For small set of short sequences, the DEA algorithm has comparable performance to Expansion Algorithm, but much better than it on many long sequences. Between the two deposition methods, the MC deposition method is slightly better than the MF method. The DEA algorithm also outperformed the Expansion Algorithm much in processing time.

To our best knowledge, our DEA algorithm is one of the best heuristic algorithms for LCS problem on large LCS instances. We believe that the use of our algorithm can facilitate researches involving the LCS problem, such as analysis on biological sequences and text sequences.

There are many other deposition and extension strategies for the LCS problem, such as look ahead strategies for deposition method, and more complicated extension method, which may lead to better performance, and we are working on these for the better heuristic results as well as the fast process. DEA algorithm on other interesting applications of the LCS problem, such as text compression and pattern recognition will also be analyzed in the future.

**Acknowledgement**

We thank Giancarlo Mauri and Gianluca Della Vedova of University of Milano-Bicocca for kindly providing us with the Expansion Algorithm source codes and relevant documentations. We also thank anonymous reviewers for insightful comments on this paper.